\documentclass[prb,a4paper,twocolumn,floatfix,showpacs,showkeys,amsmath,amssymb,nobibnotes,unsortedaddress,superscriptaddress]{revtex4-1}
\usepackage{graphicx}% Include figure files
\usepackage{pslatex}
\usepackage{xspace}
\usepackage{subfigure}
\usepackage[tight,ugly]{units}
\usepackage{array}

\usepackage{tabularx}
\usepackage{multirow}
\usepackage{booktabs}

\usepackage[pdfauthor={Jens Lorrmann},%
pdftitle={Distribution of  charge carrier transport properties in organic semiconductors with Gaussian disorder},%
pagebackref=false,
pdfkeywords={TOF, time-of-flight, transport, mobility, },
pdftex]{hyperref}
\hypersetup{
    colorlinks,%
    citecolor=black,%
    filecolor=black,%
    linkcolor=black,%
    urlcolor=black
}
\newcommand{\bfig}{\begin{figure}}
\newcommand{\efig}{\end{figure}}
\newcommand{\bit}{\begin{itemize}}
\newcommand{\eit}{\end{itemize}}

\newcommand{\tof}{\textsc{ToF}}
\newcommand{\pht}{poly(3-hexyl thiophene-2,5-diyl)\xspace}

\newcommand{\etal}{\emph{et\,al.}}

\newcommand{\dfr}[3][ ]{\frac{ {\rm d}^{#1}#2}{ {\rm d}#3^{#1}}}

\newenvironment{figref}[1]{Fig.~\ref{#1}}{}
\newenvironment{eqnref}[1]{Eqn.~(\ref{#1})}{}

\graphicspath{{figs/}{bilder/pdf/}}

\newcolumntype{M}{>{\centering\arraybackslash}m{\dimexpr.25\linewidth-2\tabcolsep}}

\begin{document}

\title{Distribution of  charge carrier transport properties in organic semiconductors with Gaussian disorder}

\author{Jens~Lorrmann}\email{jens@lorrmann.de}
\author{Manuel~Ruf}
\author{David~Vocke}
\affiliation{Experimental Physics VI, Julius Maximilian University of W{\"u}rzburg, 97074 W{\"u}rzburg, Germany}
\author{Vladimir~Dyakonov}
\affiliation{Experimental Physics VI, Julius Maximilian University of W{\"u}rzburg, 97074 W{\"u}rzburg, Germany}
\affiliation{Bavarian Center for Applied Energy Research e.V. (ZAE Bayern), 97074 W{\"u}rzburg, Germany}
\author{Carsten~Deibel}\email{deibel@disorderedmatter.eu}
\affiliation{Experimental Physics VI, Julius Maximilian University of W{\"u}rzburg, 97074 W{\"u}rzburg, Germany}

\date{\today}

\begin{abstract}

The charge carrier drift mobility in disordered semiconductors is commonly graphically extracted from time-of-flight (\tof) photocurrent transients yielding a single transit time. However, the term transit time is ambiguously defined and fails to deliver a mobility in terms of a statistical average. Here, we introduce an advanced computational procedure to evaluate \tof~transients, which allows to extract the whole distribution of transit times and mobilities from the photocurrent transient, instead of a single value. This method, extending the work of Scott~\etal~(Phys. Rev. B 46, 8603), is applicable to disordered systems with a Gaussian density of states (DOS) and its accuracy is validated using one-dimensional Monte Carlo simulations. We demonstrate the superiority of this new approach by comparing it to the common geometrical analysis of hole \tof~transients measured on \pht~(P3HT). The extracted distributions provide access to a very detailed and accurate analysis of the charge carrier transport. For instance, not only the mobility given by the mean transit time, but also the mean mobility can be calculated. Whereas the latter determines the macroscopic photocurrent, the former is relevant for an accurate determination of the energetic disorder parameter $\sigma$ within the Gaussian disorder model (GDM). $\sigma$ derived by using the common geometrical method is, as we show, underestimated instead.
%\xxx{Add scheinert2012 as reference. They fitted P3HT OFET and found+discuss sigma~0.07eV. Coehoorn+Scheinert for discussion and numbers on dark dens for kink to appear ($\hat{\sigma}^{2}==(Ef-E0)/kT \rightarrow $Boltzman fails...} 
\end{abstract}
 
\maketitle
%%
%%
%% Introduction
%%
%%   
\section{Introduction}
\label{sec:intro}
Disordered or amorphous semiconductors provide the potential of cheap future electronics with a wide variance of physical properties, e.g., the absorption spectrum. This flexibility allows to design them for specific areas of application, ranging from transistors and light emitting diodes to sunlight harvesting and converting solar cells. A common key feature determining the electrical device performance in these materials is the hopping type charge carrier transport. Therefore, during the last decades a lot of effort has been put into studies to understand the detailed transport mechanism, experimentally and theoretically.~\cite{movaghar1986, arkhipov1982, rudenko1982, lorrmann2010, strobel2010, andersson2011}

In classical descriptions the charge transport is based on the electrostatic force induced drift and on diffusion due to a spatial charge carrier density gradient. The related parameters are the charge carrier mobility $\mu$ and charge carrier diffusivity or diffusion constant $D$, respectively, which are connected by the classical Einstein~\cite{einstein1905}--Smoluchowski~\cite{vonsmoluchowski1906} relation $D/\mu = k_B T / e$. Here, $k_B$ is the Boltzmann coefficient, $e$ the elementary charge and $T$ the temperature. This concept successfully describes semiconductor devices coupling continuity and Poisson equations~\cite{wagenpfahl2010, wagenpfahl2010a} in combination with physical models concerning, e.g., dissociation~\cite{onsager1938, braun1984} or recombination.~\cite{langevin1903, thomson1896}

However, in the case of disordered systems it is known that the concept of mobility can be ill-defined~\cite{scher1975} or, in other words, the transport is determined by a broad distribution of mobilities. This fact is due to the random nature of the charge carrier transport in an energetically and spatially disordered environment. It leads to energetic relaxation of the charge carriers with time,~\cite{scher1975, nikitenko2007a, monroe1985, baranovskii2000} and to an inhomogeneous distribution of transport properties throughout thin-film devices.~\cite{yu2001, rappaport2007, bassler1993, vanderholst2009} A detailed experimental investigation of the relation between the macroscopic behaviour and the microscopic transport statistics, however, requires the extraction of transport parameter distributions. Furthermore, conclusions in terms of the energy distribution of the DOS may be drawn from the shape of these distributions. Such an approach should significantly improve the ability to evaluate the consistency between experimental data and various theoretical models. In turn, we consider it as a fundamental step towards a more general functional description of the charge carrier transport in disordered systems in terms of drift and diffusion. In several cases, such descriptions are already available,~\cite{bassler1993, schmechel2002, pasveer2005, coehoorn2005, baranovskii2000, arkhipov2001h, nikitenko2007a, scher1975} but not in the general form presented here.

First attempts to experimentally extract mobility distributions from time-of-flight~(\tof)~\cite{spear1957, kepler1960, leblanc1959} measurements or from the turn-on dynamics of polymer photo-cells where published by Scott~\etal~\cite{scott1992b} and Rappaport~\etal,~\cite{rappaport2007, rappaport2006} respectively. Unfortunately, the method by Scott~\etal~was seldom used since its publication in 1992.~\cite{bloom1997, schupper2009} The same holds for the method published by Rappaport~\etal\ Moreover the full potential utilising the statistics of the extracted parameter distributions was hardly ever discussed. In the case of Scott's method this is probably due to the fact that the method in the original form is not suitable for systems exhibiting a Gaussian density of states (DOS), found in many organic semiconductors, suh as the widely used conjugated polymer \pht (P3HT). 

In this paper we extend the method by Scott~\etal~with respect to a Gaussian DOS and reveal the advantages of this approach over the conventional geometrical technique to extract mobilities. The geometrical method determines the charge carrier mobility from a particular transit time, which is graphically extracted from \tof~transients. This transit time is defined by the intersection of the fits to the pre- and the post-transit sections of the photocurrent transient, respectively~\cite{scher1975, melnyk1993}. Henceforth we refer to the methods compared in this work as the modified Scott method and the geometric method, respectively. To verify the accuracy of the our new approach and the modifications to Scott's method we utilise a one-dimensional Monte Carlo simulation. Generally, the justification of an alternative approach is best demonstrated in comparison to a well-established standard method. Hence, the comparison is done here by evaluating the hole \tof~transients measured on P3HT with the modified Scott and the geometric method, respectively. Subsequently, we identify the statistical averages of the transit times and the mobilities from the transit time and mobility distributions, and match them against the transit times and mobilities determined from the geometric analysis of \tof~transients. 

Although convenient to use, it is, however, well known that the geometric method extracts single value transit times which account for the fastest carriers only.~\cite{marshall1987, seynhaeve1988} We demonstrate below, that the corresponding mobilities are not related to the average values and that the relative deviation depends strongly on temperature and electric field. Furthermore, the geometric method is more prone to evaluation errors, as, in the case of systems exhibiting a Gaussian DOS, the fitted regions just span over a very small part of the current transient and the regions' boundaries are judged by the eye, which may lead to huge uncertainties on a double-logarithmic scale. On the other hand, the modified Scott method, as will be shown below, needs just one fit spanning over more than one order of magnitude in time. Novikov~\etal~explicitly discussed the consequences of using poorly defined transit times or mobilities in Ref.~\onlinecite{novikov2009} and asked for a safe and reliable procedure for the analysis of \tof~data. We believe that the new approach presented here fulfils this request, by experimentally providing detailed and comprehensive information, as it is usually only known from Monte Carlo simulations. 

In Sec.~\ref{sec:expmeth}, we describe the experimental method, whereas the new approach of data evaluation is described in Sec.~\ref{sec:evalmeth}. The modeling approach is explained in Sec.~\ref{sec:simulation}. Sec.~\ref{sec:results} contains the results: In part~A we verify \eqnref{eq:scott_j0}, in part~B the new approach is used to analyse the measured \tof~transients and the findings are compared with the results obtained by the conventional geometric analysis in part~C, while in part~D the energetic disorder  parameter $\sigma$ of the P3HT hole DOS is derived.

\section{Experimental method}
\label{sec:expmeth}
The studied devices were prepared in a diode configuration, with P3HT 4002E ($\sim94\%$ regioregular; Rieke Metals Inc., without further purification) as the active material, drop cast on ITO covered glass with an aluminium electrode thermally evaporated on top. The sample thickness and excitation intensity were chosen in a way to ensure a photogeneration of charge carriers confined to the first $10\%$ of the bulk thickness. For \tof~measurements the sample was mounted in a He closed-cycle cryostat allowing field and temperature dependent studies. For the optical excitation we used the second harmonic of a pulsed \textsl{Nd:YAG} laser emitting at $\lambda=\unit[532]{nm}$, close to the absorption maximum of the polymer,~\cite{baumann2008} together with optical intensity attenuators. 

Adjusting the polarity of the applied constant electric field, one can select the type of the charge carriers being dragged through the bulk and extracted at the counter-electrode.The \tof~transients for holes in P3HT where measured in the temperatures range between $T=\unit[110 -300]{K}$ and at different electric fields between $F=\unitfrac[1.3 \times 10^{7}-1.9 \times 10^{8}]{V}{m}$.

%%
%%
%% Experimentla
%%
%%
\section{Evaluation method}
\label{sec:evalmeth}
%%
%% Tab 1
%%
\begin{table*}[th]
\centering
\caption{Overview over the four different mobility definitions used throughout this Paper.}
\small
\begin{tabular}{cccm{5cm}c|l}\toprule
&Name  &  \hspace*{.25cm}relation to transit time\hspace*{.25cm} & \hspace*{1.5cm}physical context & &\hspace*{2.25cm}Scheme\\ \midrule\midrule

\multicolumn{4}{c}{modified Scott method} &&\multirow{4}{5.5cm}{\includegraphics[scale=1]{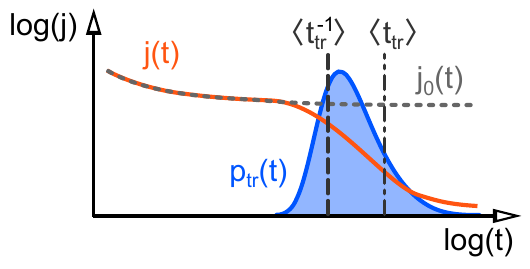}}\\[.05cm]\cmidrule{0-3}
&$\mu_{m}$ & $\propto \left\langle\dfrac{1}{t_{tr}}\right\rangle$ & ensemble average mobility; determines the macroscopic photocurrent & &\\[.55cm]
&$\mu_{tr,m}$ & $\propto \dfrac{1}{\left\langle t_{tr}\right\rangle}$ & related to average transit time; often used in Monte Carlo simulations & &\\
&  &  &  & &\\\midrule

\multicolumn{4}{c}{geometric method} &&\multirow{4}{5.5cm}{\includegraphics[scale=1]{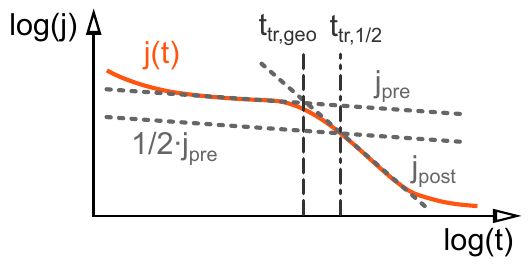}}\\[.05cm]\cmidrule{0-3}
&$\mu_{geo}$ & $\propto \dfrac{1}{ t_{tr,geo}}$ & time at which the first charge carriers reach the counterelectrode & &\\[.55cm]
&$\mu_{1/2}$ & $\propto \dfrac{1}{t_{tr,1/2}}$ & time at which $\sim50\%$ of the charge carriers were extracted & &\\
&  &  &  & &\\\bottomrule

\end{tabular}
\label{tab:mobilities}
\end{table*}
Before describing the modified Scott method, it is worth having a closer look at the photocurrent decay prior to extraction of charge carriers, as its functional approximation is an essential part of our modification.

In general the photogenerated excitons dissociate under the influence of the applied electric field, which directs the separated charge carriers to the corresponding electrodes. While one sort of charge carriers is being extracted at the illuminated electrode resulting, the other sort of charge carriers moves through the device to the counterelectrode, where they are extracted. This motion of charge carriers yields the photocurrent in the external circuit. Due to the extraction, the photocurrent steeply drops towards zero until no mobile charge carriers are left within the device. However, the photocurrent decreases even before extraction due to energetic relaxation of the charge carriers within the DOS.~\cite{monroe1985, nikitenko2007a, bassler1993, baranovskii2000} Associated with this relaxation is a dispersion of the charge carrier package being stronger than predicted for purely Gaussian (non-dispersive) transport, together with a decreasing mean drift velocity with time.~\cite{nikitenko2007a,germs2011} As a consequence, the distribution of transit times is broadened and thus the drop in the photocurrent at the extraction is blurred. 

In case of a purely exponential DOS the scenario described above can be explained within the framework of Scher and Montroll~\cite{scher1975} yielding a power law for the pre-transit ($j_0(t)\propto t^{-(1 - \alpha_{pre})}$) photocurrent. Note that this behaviour in combination with the predicted post-transit current $j(t)\propto t^{-(1 + \alpha_{post})}$ builds the basis of the geometric analysis.~\cite{melnyk1993} In contrast to the exponential DOS, where transient measurements strictly follow these power laws at least \emph{``as long as one would be able to measure''},~\cite{tessler2005} it is well known that a Gaussian DOS yields a saturation of the transient photocurrent in time, thus it is a power law with a time dependent exponent $j_0(t)\propto t^{\alpha(t)}$. The time dependence of $\alpha(t)$ is related to experimental parameters, such as the temperature and the applied electric field, and material properties (energetic and spatial disorder).~\cite{nikitenko2007a, germs2011} Unfortunately, a plain functional description of $\alpha(t)$ and, thus, the photocurrent decay $j_0(t)$, similar to the power law relation for a exponential DOS, is not available. However, taking into account that $\alpha(t)$ is a gradually decreasing function and that $\lim\limits_{t \rightarrow \infty}{\alpha(t)} = 0$, we suggest the following empirical approximation of $j_0(t)$:
\begin{subequations}
	\begin{align}  
		j_0(t) = j_{0}^{*} t^{-\alpha(t)}\\
		\alpha(t) = \left(\frac{\tau}{t}\right)^{\beta}\text{~.}
	\end{align}
	\label{eq:scott_j0}
\end{subequations}

$j_{0}^{*}$ is a scaling factor determining the photocurrent at $t =\unit[1]{s}$, while $\tau$ is a lifetime and $\beta$ is a stretching exponent defining the dynamics of the decay. We confirm the accuracy of this expression in Sec.~\ref{sec:results} using one-dimensional Monte Carlo simulations and show that all three empirically defined fitting parameters exhibit a strict temperature dependence. The fidelity of \eqnref{eq:scott_j0} can already be seen from \figref{fig:transients}.

In order to describe the full transient including extraction, the integrated number of charge carriers arriving between $t$ and ${\rm d}t$, $p_{tr}(t)$, is subtracted from $j_{0}(t)$:
\begin{equation}  
	j(t)=j_0(t)\left(1-\int_0^t p_{tr}(\hat{t}) {\rm d}\hat{t}\right) \text{~.}
	\label{eq:scott_all}
\end{equation}
Solving \eqnref{eq:scott_all} for the transit time distribution $p_{tr}(t)$ yields
\begin{equation}  
	p_{tr}(t) = -\dfr{}{t}\frac{j(t)}{j_{0}(t)}\text{~.}
	\label{eq:scott_pt}
\end{equation}
Hence, the transit time distribution $p_{tr}(t)$ can be calculated directly from the experimental photocurrent $j(t)$ and $j_{0}(t)$.

Six exemplary transit time distributions calculated from measured and simulated \tof~transients are plotted in the lower part of \figref{fig:transients}. A transit time distribution can be transformed into a mobility distribution by: $p_\mu(\mu) = \left(F t^2/d\right) p_{tr}(t)$. 

The comparison of the modified Scott and the geometric method is based on the comparison of four explicit mobility definitions. To facilitate the later discussion we shortly summarise the differences between the compared mobilities. In Tab.~\ref{tab:mobilities} we depict and list how they are obtained, we assign them to their corresponding method and put a note onto their physical relevance. In general the relation between mobility and transit time is: $\mu= L/(t_{tr}  F)$, with the sample thickness $L$ and the applied electric field $F$. Thus, the mobility is commonly determined in \tof~transients from a particular transit time $t_{tr}$. The mobility labels and the relation to the corresponding transit times can be found in Tab.~\ref{tab:mobilities} and the transit times are obtained as follows. 

In the geometrical method $t_{tr,geo}$ is identified from the intersection of two power laws fitted to the plateau ($j_{pre}$) and the trailing edge ($j_{post}$) of the photocurrent transients in double-logarithmic presentation, and $t_{tr,1/2}$ is the time at which \mbox{$1/2\times j_{pre} =j_{post}$}  (see lower scheme in  Tab.~\ref{tab:mobilities}). 

From the extracted distributions the mathematical expectations can be calculated using standard probability theory,~\cite{v.neumann1932,halmos1942} yielding the relation between the transit time $\left\langle t_{tr}\right\rangle$ and the transit time distribution $p_{tr}(t)$ as
\begin{equation}
	\left\langle t_{tr}\right\rangle= \int t_{tr} p_{tr}(t) \mathrm{d}t \mathrm{~.}
	\label{eq:ttrm}
\end{equation}
Consequently, the stochastical average of the mobility $\left\langle\mu\right\rangle$, called $\mu_{m}$ is 
\begin{align}
	\mu_{m}  &=\left\langle\mu\right\rangle= \int \mu p_{\mu}(\mu) \mathrm{d}\mu \nonumber\\
			&= \frac{L}{F}\int \frac{1}{t_{tr}} p_{tr}(t) \mathrm{d}t = \frac{L}{F}\left\langle\frac{1}{t_{tr}}\right\rangle\mathrm{~.}
	\label{eq:mum}
\end{align}

Note that generally the mathematical expectation of the mobility $\mu_{m}$ is not identical to $\mu_{tr,m} \propto \left\langle t_{tr}\right\rangle^{-1}$, as the former is proportional to the average of the inverse transit time $\left\langle t_{tr}^{-1}\right\rangle$. While $\lim\limits_{L,F \rightarrow \infty}{\mu_{m}\equiv}\mu_{tr,m}$, under normal experimental conditions at finite applied electric field $F$ and finite length $L$ of the sample under consideration, $\mu_{m}$ and $\mu_{tr,m}$ yield very different values: then, the mean mobility $\mu_{m}$ is mathematically less affected by the fraction of slow charge carriers than $\mu_{tr,m}$. The latter mobility is often found in literature as an output parameter treating the charge transport in disordered materials using Monte Carlo~\cite{bassler1993} or Master Equation~\cite{movaghar1986, pasveer2005} simulations. In case of quasi-Gaussian (non-dispersive) transport the gaussian shaped charge carrier package drifts exactly with the velocity $\mu_{tr,m} F$ through the device and $\mu_{tr,m}$ is independent of the sample thickness $L$ and the applied electric field $F$. $\mu_{m}$, instead, is linked to the average charge carrier transport and determines the macroscopic current density by $j = e n F\int \mu p_{\mu}(\mu) \mathrm{d}\mu  = e n F \mu_{m}$. 

%%
%%
%% Simulation
%%
%%
\section{Simulation method}
\label{sec:simulation}
  
In order to verify our approach, a one-dimensional Monte Carlo simulation~\cite{silver1971,marshall1977} was implemented to study charge carrier transport by hopping. It considers the capture and the emission of charge carriers according to the multiple trapping and release (MTR) model. Before further specifying the simulation procedure, we first recall some details of the MTR-model developed by Schmidlin~\etal~\cite{schmidlin1977} and Noolandi~\etal~\cite{noolandi1977} 

Originally the presence of extended states in the valence and conduction bands and localised trapping states in the band gap were considered. In general to model the charge carrier transport any localised state in the band gap needs to be considered as a possible target for a charge carrier jump. This, e.g. by kinetic Monte Carlo simulation in three dimensions, requires considerable amounts of both computer memory and computation time. However, transport within the band gap is very unlikely due to the fact that the transfer rate between two states depends exponentially on the electronic coupling, which is very small in between states in the band gap. Thus, a charge carrier in the band gap will most probably be thermally excited to the mobility edge, which separates the localised states with a low density from the extended states with a high density. The MTR model replaces the full picture of hopping by a quasi-free transport until trapping of charge carriers above the mobility edge and a thermal emission back to extended states of charge carriers below. 

In disordered organic semiconductors instead, the valence and conduction bands are described in terms of a Gaussian distribution of localised states. Transport in these localised states occurs via hopping. The hopping transport resembles the MTR process near and below a temperature dependent transport energy $E_{trans}$.~\cite{grunewald1979} Hence, in such a system, the transport energy $E_{trans}$ plays the role of the mobility edge.~\cite{monroe1985, baranovskii1997}

In our simulation we model the transport of noninteracting charge carriers. A charge carrier is either transfered to the transport energy $E_{trans}$ by thermal excitation from a trap state or moves quasi-free above the transport energy $E_{trans}$ until it is trapped again or reaches an electrode and is extracted. By integrating the duration of each individual process the model gets a time dependence. This allows the simulation of transient measurement techniques such as \tof. 

%%
%% Tab 2
%%
\begin{table}[b]
\centering
\caption{Simulation parameters. The parameters were determined by reconstruction of experimental \tof~transients.}
\small
\begin{tabular}{c*{4}{@{\extracolsep{6mm}}c}}\toprule
Parameter  &  Value & Unit & Source\\ \midrule
$\sigma$ & $69.9$ & meV & see Text\\ 
$\nu_0$ & $1\times 10^{13}$ & s$^{-1}$ & Ref.~\onlinecite{deibel2009a}\\ 
$E_{trans}$ & $0.0$ & eV & Approx.\\
$\gamma$ & $3.91\times 10^{9}$ & m$^{-1}$ & Fit \\ 
$a_x$ & $1.26\times 10^{-9}$ & m & Fit \\
$\mu_0(F,T)$ & $(0.8 - 2.15)\times 10^{-8}$ & m$^2$(Vs)$^{-1}$ & Fit \\\bottomrule
\end{tabular}
\label{tab:MCParams}
\end{table}

Initially, all charge carriers are set in vicinity to the illuminated electrode and reside at the transport energy $E_{trans}$. During the time until capture
\begin{equation}
	\tau_{c} = \frac{\ln{\left(\zeta\right)}}{\tau_{0}} 
	\label{eq:capture_time}
\end{equation}
the charge carriers at the transport energy move quasi-free with the mobility $\mu_0(F,T)$ for a distance $\Delta x= \tau_c\times \mu_0(F,T)\times F$. Where $\tau_{0} = \left[\nu_0\times \exp{( -2 \times \gamma \times a_x)}\right]^{-1}$ is the trapping rate with the attempt-to-escape frequency $\nu_{0}$, the inverse localisation length $\gamma$ and the intermolecular distance $a_x$. $\zeta$ is a uniformly distributed random number between $]0,1]$, $F$ is the electric field and $T$ the temperature. The charge carrier is then randomly trapped into one of the trapping levels $E_{trap}$ with a probability according to the Gaussian energy distribution of these levels. The release time from a trap level is given by
\begin{equation}
	\tau_{r} = \tau_{0} \exp{\left( \frac{E_{trans} - E_{trap} - a_x F}{k_B T}\right)} \ln{\left(\zeta\right)} \ ,
	\label{eq:sim_releasetime}
\end{equation}
where $k_B$ the Boltzmann constant. To accurately fit the field dependence of our \tof~transients we included an exponential field reduction factor $a_x F/(k_B T)$ in \eqnref{eq:sim_releasetime} accounting for a field-induced detrapping.~\cite{cottaar2010, miller1960, schubert2013} This factor yields the well known Poole-Frenkel effect~\cite{frenkel1938} by effectively reducing the hopping barrier for a charge carrier in direction of the electric field.~\cite{dunlap1999} Tyutnev~\etal~implement an equivalent modification in their \emph{MTRg}-model~\cite{tyutnev2012} to consistently explain the Pool-Frenkel effect in accordance with three-dimensional kinetic Monte Carlo simulations.~\cite{deibel2009a, strobel2010, bassler1993}.

From the temporal evolution of the charge carriers' positions in the sample, a mean current can be calculated by $j(t) = e  n_0  v(t)$ where $e$ is the elementary charge, $n_0$ the number of initially generated carriers and $v(t)$ the carriers' mean velocity.

By suitable parameterisation of the model, experimental measurements were reconstructed by simulation, offering a deeper insight into microscopic charge transport phenomena. To decrease the number free fitting parameters we firstly fixed a couple of the parameters to experimentally determined values as well as physically reasonable literature values. 

The energetic disorder parameter was set to be $\sigma=\unit[69.9]{meV}$. This value was calculated from the temperature dependence of the zero-field mobility in terms of the Gaussian disorder model.~\cite{bassler1993} The attempt-to-escape frequency was chosen equal to $\nu_0=1\times 10^{13} {\rm~s^{-1}}$ from Ref.~\onlinecite{deibel2009a}. We assumed a fixed transport energy at $E_{trans}=\unit[0]{eV}$. Fixed, because we found just little impact of a temperature dependent transport energy for the temperature range considered and equal to $\unit[0]{eV}$, because the effect of the real transport energy value on the transient shape is implicitly considered in the two free fitting parameters $a_x$ and $\mu_0(F,T)$. In accordance with our experience Germs~\etal\ found a temperature dependence of the transport energy comparable to the fitting uncertainty for $\sigma/(k_B  T)<5.1$,~\cite{germs2011} --- this is in our case $T>\unit[160]{K}$. Further input parameters are the inverse localisation length $\gamma$ and the intermolecular distance $a_x$ and the charge carrier mobility $\mu_0(F,T)$. 

To fully calibrate the simulation we simultaneously fitted a set of measurements at different temperatures between $T=\unit[175 -300]{K}$ and at different electric fields $F=\unitfrac[1.3\times 10^{7}-1.9\times 10^{8}]{V}{m}$ using the global minimisation algorithm of differential evolution.~\cite{storn1997} The used parameters are listed in Tab.~\ref{tab:MCParams}, where we explicitly mark a parameter as fixed or fitted.

Note that from the one-dimensional Monte Carlo simulations the transit time distribution $p_{tr}(t)$ can be calculated directly by the monitored gradual decrease of the number of charge carriers inside the device $N(t)$. The relation is
\begin{equation}
  p_{tr}(t) = -\frac{dN(t)}{dt} \ .
 \label{eq:pt_fromsim}
\end{equation}

%%
%%
%% Results
%%
%%
\section{Results}
\label{sec:results}
\subsection{Verifying the initial photocurrent approximation \eqnref{eq:scott_j0}}
\label{subsec:verify}

\begin{figure}[t]
    \centering
    \includegraphics[width=.5\textwidth]{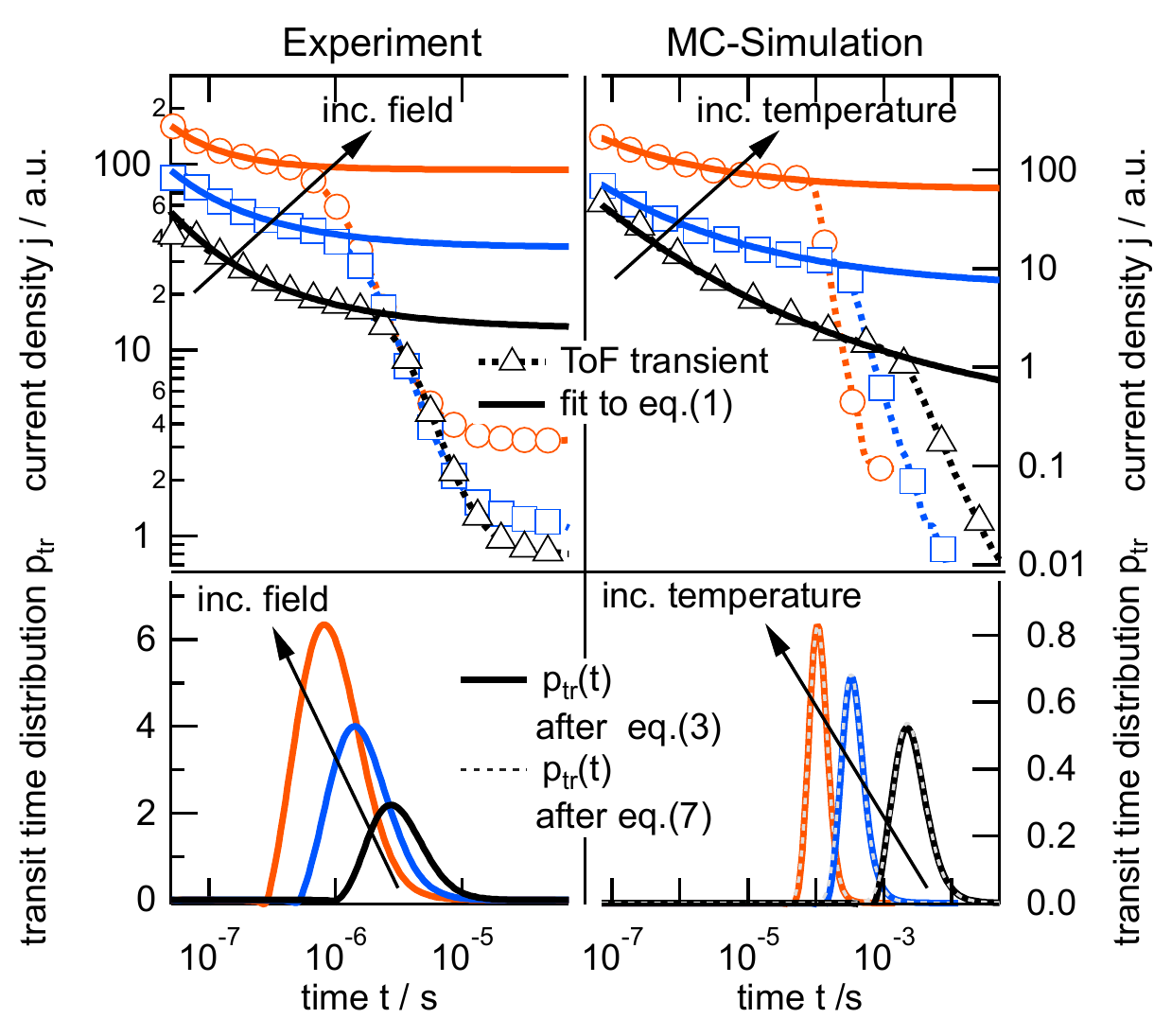}\hfill
    \caption{\textbf{Upper part}: Measured \tof~transients of hole currents in P3HT at $T = \unit[260]{K}$ for different applied voltages $F = \unitfrac[6.7\times 10^{7}]{V}{m},\ \unitfrac[1.1\times 10^{8}]{V}{m},\ \unitfrac[1.5\times 10^{8}]{V}{m}$~(left). Simulated \tof~transients $F = \unitfrac[1\times10^{7}]{V}{m}$ for different temperatures $T = \unit[300]{K},\ \unit[250]{K},\ \unit[200]{K}$~(right). The pre-transit decay of the photocurrent (dotted lines with markers) is reproduced by the fit to \eqnref{eq:scott_j0} (lines without markers). \textbf{Lower part}: Transit-time distribution $p_{tr}(t)$ for the experimental (left) and the simulated (right) transients calculated from \eqnref{eq:scott_pt}. Note that the extracted distributions (solid lines, right) match exactly the distributions (dotted lines, right) calculated by \eqnref{eq:pt_fromsim} from the temporal decrease of the amount $N$ of charge carriers inside the device.}
    \label{fig:transients}
\end{figure}

At first we demonstrate the quality of the empirical expression~\eqnref{eq:scott_j0} of $j_0(t)$. To this end, we simulated a set of temperature dependent ($T=\unit[200]{K} -\unit[300]{K}$) \tof~transients using our one-dimensional Monte Carlo simulation and compare the transit time distributions obtained by the modified Scott method with the temporal decrease of charge carriers inside the device according to \eqnref{eq:pt_fromsim}. The parameterised values used in the simulation are listed in Tab.~\ref{tab:MCParams}. In the upper right part of \figref{fig:transients} the fit of \eqnref{eq:scott_j0} to the simulated pre-transit photocurrent decay is plotted together with the full simulated transients. The approximation of the photocurrent by $j_0(t)$ is excellent over up to four orders of magnitude in time. The equivalence of the extracted distributions is shown in the lower right part of \figref{fig:transients}: both transit time distributions match perfectly well over the whole studied temperature range and verify the validity of the extrapolation of $j_0(t)$ to post-transit times. The corresponding average transit times of the simulated ensemble of charge carriers and the values extracted from the transit time distributions $p_{tr}(t)$ by \eqnref{eq:ttrm} are equal within the range of numerical precision ($\Delta \left\langle t_{tr}\right\rangle/\left\langle t_{tr}\right\rangle < 0.5\%$). 
Further testing of \eqnref{eq:scott_j0} is provided in Appendix~\ref{AppA}.

Based on these results, we conclude that \eqnref{eq:scott_j0} is a suitable parametric approximation for the photocurrent decay and that, thus, the modified Scott method can be applied to systems with a Gaussian DOS.
%%
%% Fig 1
%%

\subsection{Evaluating hole photocurrents with the new approach}
\label{subsec:newapproach}

We examined \tof~transients measured on P3HT utilising the presented approach. The findings are compared with the results obtained by the conventional geometric analysis in Sec.~\ref{subsec:comparison}. In the upper left part of \figref{fig:transients} three \tof-transients, parametric in the applied electric field at $T=\unit[260]{K}$, are plotted together with fits of $j_{0}(t)$ according to \eqnref{eq:scott_j0} to the initial decay of the experimental photocurrent $j(t)$. The excellent agreement between the data and the $j_{0}(t)$ over almost two orders of magnitude in time was achieved for all measured temperatures and electric fields. However, it is expected that at very short times $j_0(t)$ is not suitable, since then the current decay $j(t)$ on the one hand involves a contribution of the quickly extracted electrons at the illuminated contact superimposing the hole conduction current and on the other hand is limited by the setup's $RC$ time. The respective transit time distributions, plotted in the lower left part of \figref{fig:transients}, were calculated from the experimental current density $j(t)$ as well as the fitted decay $j_{0}(t)$ according to \eqnref{eq:scott_pt}. These transit time distributions --- and consequently the mobility distributions --- are approximately inverse Gaussian~\cite{schroedinger1915,tweedie1945,siegert1951} distributed, in good agreement with results obtained from measurements on poly(2-methoxy,5-(2'-ethyl-hexoxy)-p-phenylene vinylene) (MEH-PPV) thin films performed by Rappaport~\etal~\cite{rappaport2007, rappaport2006} and on diketopyrrolopyrrole-naphthalene copolymer (PDPP-TNT) thin film transistors by Ha~\etal\cite{ha2012} 
%%
%% Fig 2
%%
\begin{figure}[t]
    \centering
    \includegraphics[width=.45\textwidth]{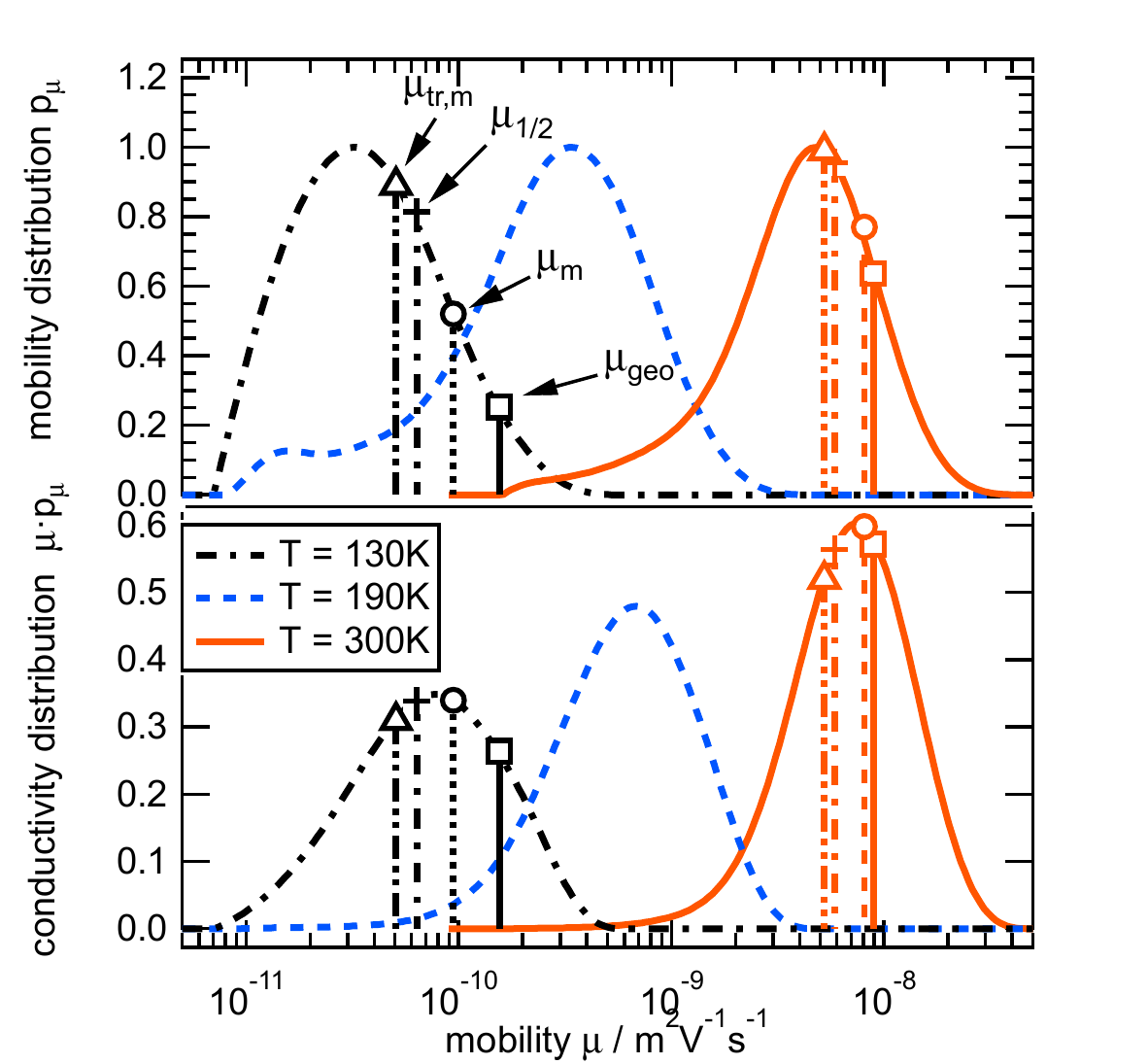}\hfill
    \caption{\textbf{Upper part}: Mobility distributions $p_{\mu}(\mu)$ for three different temperatures $T=\unit[130, 190, 300]{K}$ (from left to right) at a fixed electric field of. $F=\unitfrac[1.2\times 10^{8}]{V}{m}$. The vertical straight lines illustrate the four different definitions of mobility $\mu_{tr,m}$ (dash-double-dotted), $\mu_{m}$ (dashed), $\mu_{geo}$ (solid) and $\mu_{1/2}$ (dash-dotted) for $T=\unit[130]{K}$ (left) and  $T=\unit[300]{K}$ (right). \textbf{Lower part}: Conductivity distributions $\mu p_{\mu}(\mu)$ corresponding to the distributions above.}
    \label{fig:fig3}
\end{figure}

Interesting to note is that inverse Gaussian distributions are well known to describe the \emph{``First Passage Time Probability''}~\cite{tweedie1945,siegert1951} since Schr\"odinger's theoretical work about brownian motion of particles under a constant force causing drift.~\cite{schroedinger1915} This \emph{``First Passage Time Probability''} is actually equivalent to the distribution of transit times if a non-dispersive Gaussian charge carrier transport, which is nothing else than a Brown motion with drift, is assumed.

Derived mobility distributions $p_{\mu}(\mu)$ are displayed in the upper part of \figref{fig:fig3} for three different temperatures at a fixed electric field. Note the different asymmetric shape of $p_{\mu}(\mu)$ at different temperatures: For the high temperature ($T=\unit[260]{K}$), the maximum of the distribution is shifted to higher mobilities, according to many fast charge carriers, whereas the left edge is slowly decreasing towards slower mobilities. For the intermediate temperature ($T=\unit[190]{K}$), the shape is quite symmetric, but broader, with an additional long tail of slow charge carriers. At low temperature ($T=\unit[140]{K}$) the distribution's maximum shifts to lower mobilities and the width is further broadened.

A more intuitive representation of the extracted distributions is given by the conductivity distribution $\mu p_\mu(\mu)$ plotted in the lower part of \figref{fig:fig3}. It shows the contribution of each mobility to the total current density $j_t$. From \figref{fig:fig3} it can be seen that, although an enhanced number of low mobility charge carriers show up in the left edge of the mobility distribution at $T=\unit[140]{K}$, their contribution to the total current density $j_t$ is small as seen from the conductivity distribution in the graph \figref{fig:fig3} below. Actually the main contribution comes from charge carriers with a mobility around $\mu_{m}$. We point out that, using the transport energy definition from Schmechel,~\cite{schmechel2002} $\mu_{m}$ is the mobility of charge carriers around the transport energy.

\subsection{Comparison of the new approach with the common geometrical method}
\label{subsec:comparison}
Next we reveal the advantage and accuracy of the modified Scott method by comparing the results of our data evaluation on experimental data with results obtained with the well-established geometric approach. Therefore, we remind the reader of the two different mobility definitions for each of the two approaches listed in Tab.~\ref{tab:mobilities}.

The four different mobilities are compared in \figref{fig:fig3} for two temperatures. For both temperatures $\mu_{geo}$ has the highest value followed by $\mu_{m}$ and $\mu_{tr,m}$ is further shifted to lower mobilities. $\mu_{1/2}$ is always situated between $\mu_{m}$ and $\mu_{tr,m}$. For the two cases shown in \figref{fig:fig3} the relative error of the geometric mobility to the mean mobility is $\sim 0.10$ at $T=\unit[300]{K}$; it strongly increases at $T=\unit[130]{K}$ to $\sim0.64$.

\begin{figure}[t]
    \centering
    \includegraphics[width=.45\textwidth]{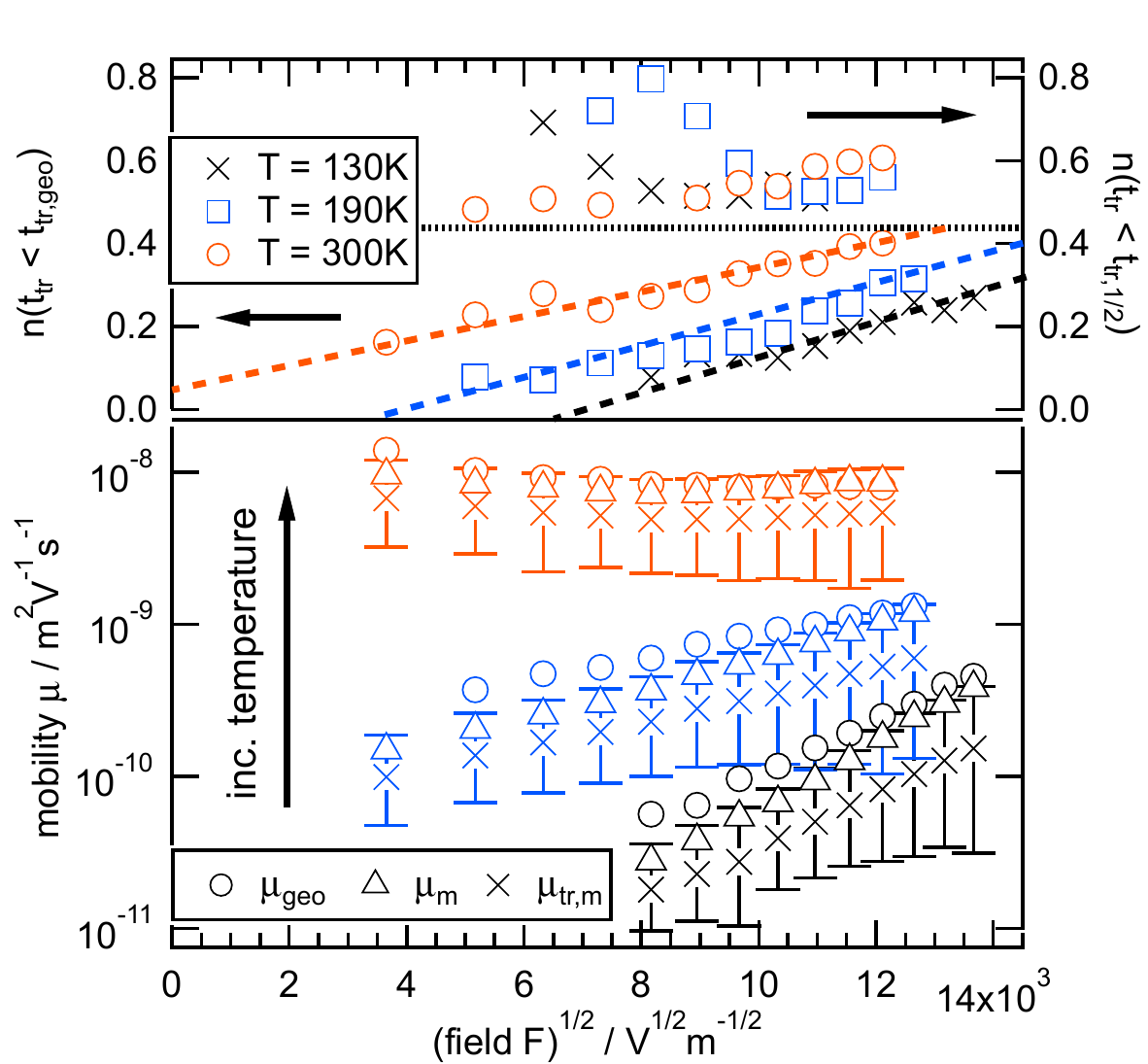}\hfill
    \caption{\textbf{Upper part, Left axis}: Fraction of charge carriers, which are extracted before the geometric transit time $t_{tr,geo}$, to all charge carriers extracted. A temperature and field dependent deviation is seen by the straight lines, which were added as a guide to the eye. \textbf{Upper part, Right axis}: Fraction of charge carriers, which are extracted before $t_{tr,1/2}$, to all charge carriers extracted. \textbf{Lower part}: Temperature and field dependence of differently defined charge carrier mobilities. As the geometric transit time only accounts for the fastest charge carriers, the mobility at the geometric transit time $\mu_{geo}$ lies at systematically higher values. The mobility $\mu_{tr,m}$ from the mean transit time together with its standard deviation is indicated by the error bars. The mean mobility $\mu_{m}$ is physically most relevant for the description of an experiment.}
    \label{fig:fig4}
\end{figure}

In order to quantify the deviation of the geometric transit time from the physically more relevant transit time distribution, the normalised density of charge carriers extracted until the transit time $t_{tr,geo}$ is studied in the upper part of \figref{fig:fig4}. While the geometric method accounts for the fastest $40\%$ of charge carriers at room temperature and high fields, this value drops to below $5\%$ at lower temperatures and fields. The field independent relation $t_{tr,geo}=A(T,\sigma) t_{tr,m}$ between the average transit time $t_{tr,m}$, calculated by averaging the transit time distribution $p_{tr}(t)$, and the geometric transit time $t_{tr,geo}$ found by Freire~\etal~\cite{freire2003}  can not be confirmed from our data. %This can be seen from the dashed lines in the upper part of \figref{fig:fig4} showing different slopes.

 Long since it is known that $\mu_{geo}$ accounts only for the fastest charge carriers~\cite{marshall1987, seynhaeve1988} and, thus, it was often recommended to use $\mu_{1/2}$ instead.~\cite{seynhaeve1988} However, the normalised density of charge carriers extracted until $t_{tr,1/2}$ shows also a slight field dependence at room temperature varying between $50\%-60\%$. A strong negative field dependence is instead detected for lower temperatures. However, at high electric fields nearly no temperature dependence is obtained and $\sim 60\%$ of the charge carriers were extracted until $t_{tr,1/2}$. 
%%
%% Fig 4
%%

As the actual charge transport is determined by the mean mobility, a systematic field and temperature dependent error is made when relying on the geometric method. Although, the mobility values determined from $t_{tr,1/2}$ better represent the mean values, for very dispersive transients with a bad signal-to-noise ratio $t_{tr,1/2}$ can not be obtained from experimental data, while a transit time distribution can still be extracted.

A further advantage of the modified Scott method is that besides the mathematical expectations the mobility and transit time can be provided with their standard deviation as a measure of the transport dispersion. In the lower part of \figref{fig:fig4} we compare the field dependence of the different mobilities $\mu_{geo}$, $\mu_{m}$ and $\mu_{tr,m}$ ($\mu_{1/2}$ is left out for clarity). Additionally we plotted the standard deviations of the mean transit time $t_{tr,m}$ accounting for the transit times shorter than $t_{tr,m}$ and for the longer transit times, respectively. As in \figref{fig:fig3}, \figref{fig:fig4} confirms that the geometric mobility $\mu_{geom}$ always exceeds both $\mu_{m}$ and $\mu_{tr,m}$. Furthermore, the geometric mobility $\mu_{geom}$ hardly lies within the bounds of the mean transit time mobilities' standard deviation. Only at high electric fields and high temperatures this is the case. It can be nicely seen that the standard deviation of the short transit times in \figref{fig:fig4} shows a negligible field, but strong temperature dependence, as in Ref.~\onlinecite{bassler1993}. Yet, the electric field and the temperature strongly increase the standard deviation for the long transit times. We attribute this strong asymmetry of the transit time distribution $p_{tr}(t)$ around $t_{tr,m}$ to field assisted diffusion.~\cite{arkhipov1982,nenashev2010a, nenashev2010b} A detailed investigation of this phenomena will be published elsewhere.

\subsection{Extracting an accurate energetic disorder parameter $\sigma$ for hole transport in P3HT}
\label{subsec:sigma}

%In general decreasing temperature and electric field enhance this deviation, yielding a wrong temperature and electric field dependence of the charge carrier drift mobility, as can be seen from the different slopes of the zero field mobilities in \figref{fig:fig5}. Long since it is known that $\mu_{geo}$ accounts only for the fastest charge carriers~\cite{marshall1987, seynhaeve1988} and thus it was often recommended to use $\mu_{1/2}$ instead.~\cite{seynhaeve1988} However, we found neither a compliance of $\mu_{1/2}$ with $\mu_{m}$ nor with $\mu_{tr,m}$.

% and that, by means of common theoretical models~\cite{brinza2006a}, the mobility axis can thus be transformed into an energy axis. Hence, the distributions comprise informations about the density of occupied states and by varying temperature and electric field it is possible to map a wide range of the DOS the charge carrier transport takes place in~\cite{longeaud2007}. But this topic is out of the scope of this paper. 

\begin{figure}[t]
    \centering
    \includegraphics[width=.45\textwidth]{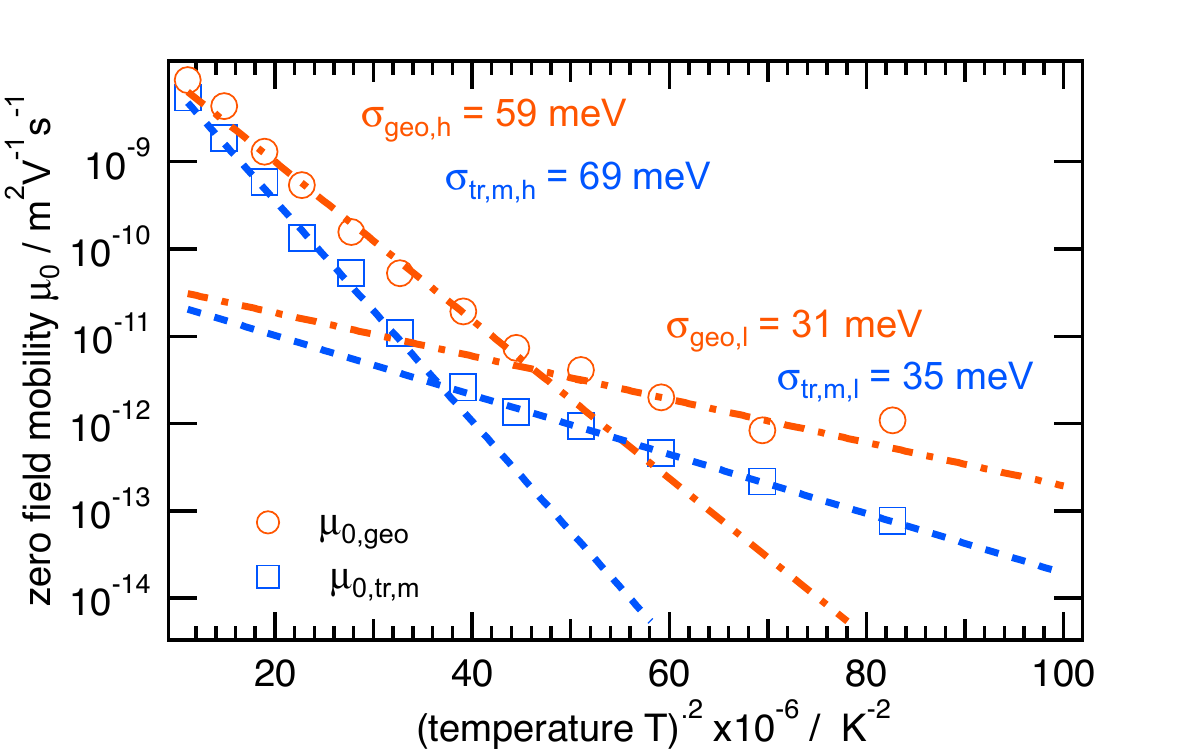}\hfill
    \caption{Zero field mobilities for  geometric $\mu_{0,geo}$ and mean transit time mobilities $\mu_{0,tr,m}$. The temperature dependent error of the geometric mobility can be clearly seen and results in an underestimation of the extracted energetic disorder $\sigma$of the DOS. The transition to a weaker temperature dependence at low temperatures is more attributed to a violation of the common Boltzmann approximation than to the transition from dispersive to non-dispersive transport. }
    \label{fig:fig5}
\end{figure}

Finally, in terms of the framework of the GDM,~\cite{bassler1993} we can make use of the extracted temperature and field dependent \tof~mobilities to gain the characteristic energetic disorder $\sigma$ of the DOS. B\"assler~\etal~used the mobility given by the mean transit time $\mu_{tr,m}$ for the derivation of their famous parametric GDM mobility equation [Eq.~(15) in Ref.~\onlinecite{bassler1993}]. Consequently, as our new approach allows for the extraction of this quantity, we can calculate an accurate $\sigma$ for the hole transport in P3HT. To this end, we demonstrate in \figref{fig:fig5} the different scaling of the zero field geometric mobility $\mu_{0,geo}$ and the zero field mean transit time mobility $\mu_0(t_{tr,m})$ with temperature. The extracted energetic disorder $\sigma_{geo,h}=\unit[59]{meV}$ is significantly lower by about $14\%$ for the geometric mobilities compared to $\sigma_{m,h}=\unit[69]{meV}$ calculated from $\mu_{tr,m}$. The found systematic error of the geometric method leads, thus, to a wrong estimate of the energetic disorder parameter $\sigma$. Comparable values for $\sigma_{geo,h}$ were found by Mauer~\etal~\cite{mauer2010} using the geometrical method. For P3HT Sepiolid P100 and Sepiolid P200 (BASF SE) $\sigma=\unit[56]{meV}$ and $\sigma=\unit[58]{meV}$ was obtained, respectively. However, our results suggest that these values are underestimated due to the used method of evaluation. From our findings we expect the real energetic disorder to be closer to $\sigma\approx\unit[64]{meV}$ for the hole transport in Sepiolid P100 and $\sigma\approx\unit[66]{meV}$ for Sepiolid P200. An energetic disorder $\sigma_{h}=\unit[70]{meV}$, in accordance with our results, is found as a best fit parameter by Scheinert \etal~\cite{scheinert2012} fitting organic transistor transfer and output characteristics with a two dimensional numerical device simulation. This simulation takes a Gaussian density of states into account, equivalent to our one dimension Monte Carlo simulation, which correctly approximates our experimental photocurrent transients with varying temperature assuming $\sigma_{h}=\unit[69]{meV}$. Thus, by using the distribution of transit times, as described above, very reliable parameters can be extracted. 

%%
%% Fig 5
%%
Besides the impact of the temperature and field dependent error of $\mu_{0, geo}$, its deviation to the mean mobility $\mu_{m}$ can also be understood regarding the energetic distribution of the charge carriers. While the fastest charge carriers just hop via states around the transport energy and thus ``feel'' a smaller energetic disorder, the mobility calculated from the distribution of all charge carriers accounts for all states in the DOS. 

Note that, besides the extracted statistical parameters discussed herein, the mobility and transit time distributions involve much more information about the energy distribution of the DOS. At temperatures below $T\approx\unit[190]{K}$ the slope of the zero field mobilities in \figref{fig:fig5} changes and become smaller. The slope of the geometric values is also lower and the relative difference is $11\%$. Albeit the evaluated values of $\sigma_{geo,l}=\unit[31]{meV}$ and $\sigma_{m,l}=\unit[35]{meV}$ are not related to the energetic disorder $\sigma$ of the DOS and are just given to number the different slopes. Furthermore, we do not necessarily relate this effect to a transition from dispersive to non-dispersive transport, as commonly suggested. First of all, we do not see a significant change in the photocurrent shape around $T\approx\unit[190]{K}$ and second we observe the very same temperature behaviour of the mobility from steady-state field effect measurements~\cite{winter2009} on the same material system. As, in the case of the modified Scott method, we can rule out a systematic error due to the evaluation procedure as the nature of this behaviour, we attribute this more to a violation of the common Boltzmann approximation utilised in the GDM, which predicts the constant slope of $\ln(\mu)\ vs\ 1/T^2$. Instead, at low temperatures, when the effective energetic disorder $\sigma^2/(k_B T)$ is equal or less than the Fermi energy $E_f$, the full Fermi-statistics need to be taken into account considering the charge carrier density effects on the mobility.~\cite{schmechel2002, coehoorn2005,scheinert2012,paasch2010} 

%%
%%
%% Conclusion
%%
%%
\section{Conclusion}

We have introduced the modified Scott method to derive the charge carrier mobility and transit time distribution in energetically disordered semiconductors exhibiting a Gaussian DOS from photocurrent transients. 

A prerequisite of this new approach, which is based on the earlier work of Scott~\etal,\cite{scott1992b} is the functional description of the photocurrent decay in the pre-transit regime and the extrapolation into the post-transit regime. However, it is known that the initial photocurrent decay in these systems cannot be described by a plain analytical relation,~\cite{nikitenko2007a, germs2011} such as the power law decay in systems with an exponential distribution of localised states within the band gap.~\cite{scher1975} Here, we proposed an empirical approximation [\eqnref{eq:scott_j0}] for the pre-transit photocurrent, which accurately fits our experimental data measured on P3HT. The validity of this functional approximation and the reliability of its extrapolation were shown by one-dimensional Monte Carlo simulations. The transit time distributions, which we either determined by evaluating simulated photocurrents with the new method or directly from the temporal decrease of the amount of charge carriers inside the simulated device, match exactly.

By applying this method to hole \tof~transients measured on P3HT over a wide temperature and electric field range we demonstrated its convenient usage and the superiority compared to the commonly used geometric method. In contrast to the latter the new approach has the following advantages:
\begin{itemize}
	\item it is able to extract parameter distributions instead of a single value only, providing detailed and comprehensive information, as it is usually only known from Monte Carlo simulations
	\item the extracted distributions provide detailed information about the transporting medium, such as, e.g., a statistical mean mobility determining the macroscopic photocurrent or the dispersion of mobilities, exhibiting much potential for further careful studies and input for theoretical considerations
	\item it is less error prone, as the pre-transit fit spans over more than one order of magnitude in time and the extracted stochastical values are numerically averaged 
	%\item straight forward application allows for full automatic numerical evaluation of photocurrent transients
\end{itemize}
Furthermore, we found that the geometric method clearly overestimates the mobility defined in terms of a mean drift, which is the expectation value $\mu_{m}$ of the mobility distribution. This deviation becomes even more prominent at lower electric fields and temperatures. The analysis of hole photocurrents with the new approach revealed an accurate energetic disorder parameter $\sigma_h=69~{\rm meV}$ for P3HT. This value was calculated from the mobility given by the mean transit time $\mu_{tr,m}$, as this is the relevant parameter considered in the framework of the GDM.~\cite{bassler1993} By using the less accurate geometrical method we determined a $\sim 14\%$ smaller value of $\sigma_{h,geo}=59~{\rm meV}$ instead.

\section*{Acknowledgments}

The current work is supported by the Deutsche Forschungsgemeinschaft (DFG), project EiNDORSE (DE~830/9-1). C.D. gratefully acknowledges the support of the Bavarian Academy of Sciences and Humanities.

\appendix
\section{Temperature scaling of Eqn. (1)}
\label{AppA}

%%
%% Fig A1
%%
\begin{figure}[t]
	\centering
	\includegraphics[width=0.5\textwidth]{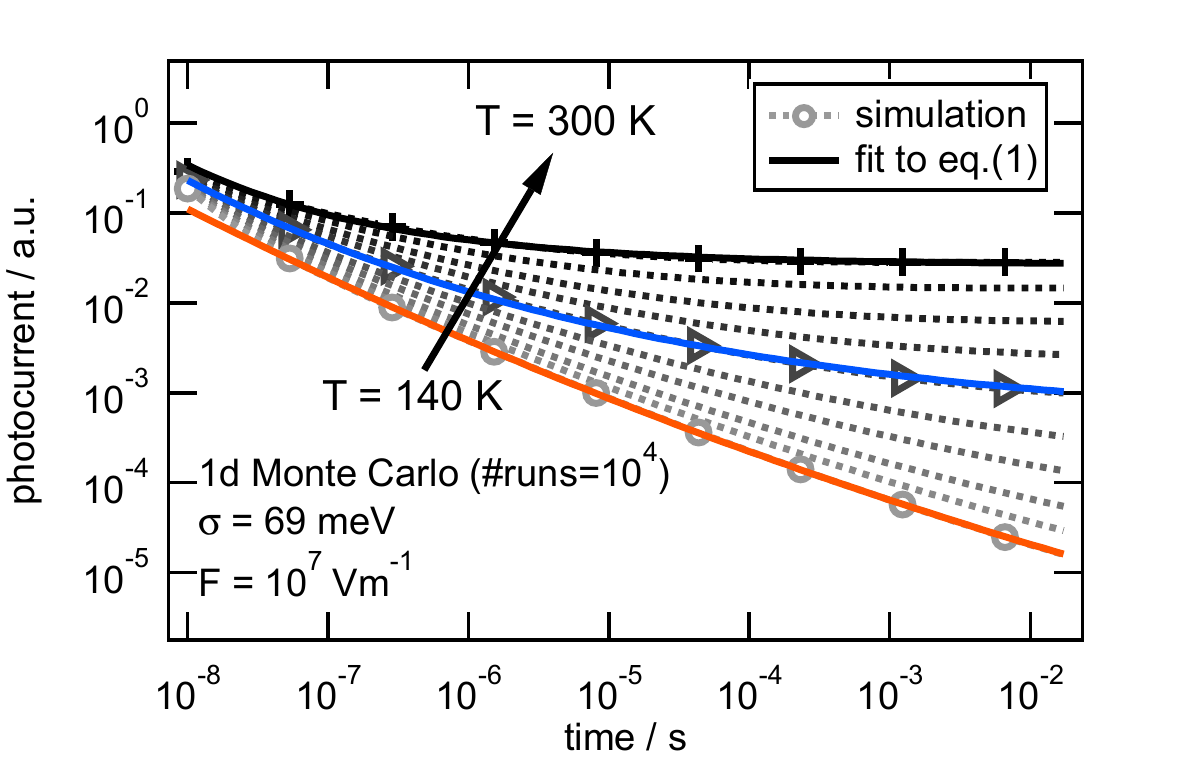}\hfill
 	\caption{Simulated photocurrent transients neglecting extraction (dotted lines and symbols) together with three exemplary fits to \eqnref{eq:scott_j0} (straight line). The simulation temperature varies between $T=\unit[140]{K}$ ($\bigcirc$) via  $T=\unit[210]{K}$ ($\rhd$) to $T=\unit[300]{K}$ ($+$) and the parameters from Tab.~\ref{tab:MCParams} were used.}
	\label{fig:1dMC_decay}
\end{figure}

In \figref{fig:1dMC_decay} photocurrent decays are shown, which where simulated with the one-dimensional Monte Carlo simulation assuming an infinite sample, such that no charge carrier extraction occurs. The same parameter set listed in Tab.~\ref{tab:MCParams} was used and we fitted the resulting transients with \eqnref{eq:scott_j0}. Three exemplary fits are plotted in \figref{fig:1dMC_decay}, too, and \eqnref{eq:scott_j0} approximates the initial photocurrent very well even down to $T=\unit[140]{K}$. In fact \eqnref{eq:scott_j0} still works for $T<\unit[140]{K}$. With the used simulation parameters, the initial decay curvature  for $T<\unit[140]{K}$ does not change much anymore approaching more or less a straight line in \figref{fig:1dMC_decay}.

%%
%% Fig A2
%%
\begin{figure}[t]
	\centering
	\includegraphics[width=0.5\textwidth]{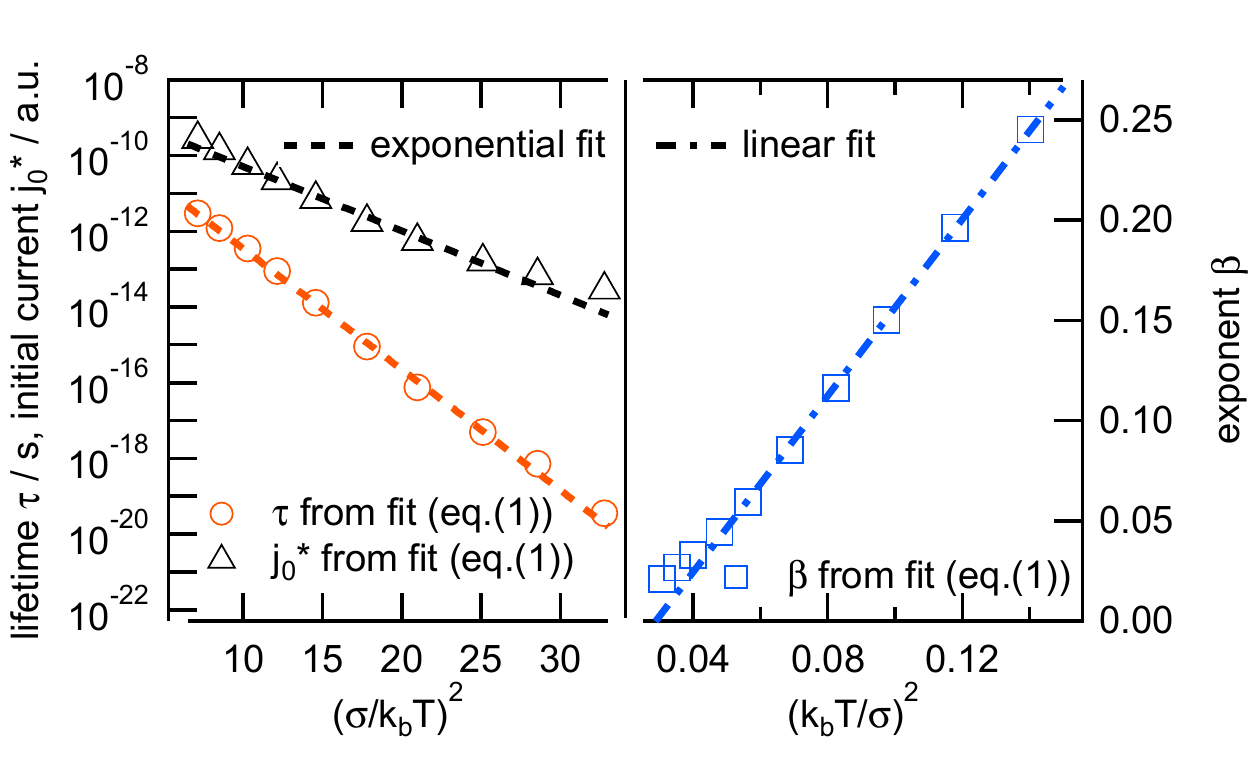}\hfill
 	\caption{Temperature dependence of the empirical fitting parameters $j_{0}^{*}$, $\tau$ and $\beta$ together with exponential and linear fits. $\log(j_{0}^{*})$ and $\log(\tau)$ are proportional to $1/T^{2}$, while $\beta$ scales linearly with $T^{2}$.}
	\label{fig:fit_params}
\end{figure}

The mathematical meaning of three fitting parameters is the following: $\tau$ defines the temporal current saturation. $j_0^{*}$ fixes the current value of \eqnref{eq:scott_j0} at the two points in time $t=\tau$ and $t=\unit[1]{s}$ as  $j_0(t=\tau)=j_0^{*} \tau$ and $j_0(t=\unit[1]{s}) =j_0^{*}$. Finally, $\beta$ specifies the curvature between these two fixed points in time --- the larger $\beta$ the faster the decay. Thus, as relaxation is accelerated with increasing temperature, $\beta$ increases with temperature.  Without claiming any real physical meaning of the three fitting parameters, further evidence for the validity of \eqnref{eq:scott_j0} comes from the fact that although \eqnref{eq:scott_j0} is an empirical approximation $j_0^{*}$, $\tau$ and $\beta$ follow a strict temperature dependence. This can be seen in \figref{fig:fit_params}. Within a temperature range from $T=\unit[140]{K} -\unit[300]{K}$ we found that $j_0^{*}$ and the lifetime $\tau$ scale as 
\begin{align*}
	j_0^{*} \propto \exp{\left[-0.4\left(\frac{\sigma}{k_B T}\right)^{2}\right]}&\\
	\tau \propto\exp{\left[-0.74\left(\frac{\sigma}{k_B T}\right)^{2}\right]}& \ ,\\
\end{align*}
respectively. The exponent $\beta$ depends on $\left(\frac{k_B  T}{\sigma}\right)^{2}$ as
\begin{equation*}
	\beta = 2.20 \left(\frac{k_B  T}{\sigma}\right)^{2}-0.063\mathrm{.}
\end{equation*}

%Due to the observed temperature scalings in the simulated temperature range in \figref{fig:fit_params}, $\left(\sigma/(k_b\times T)\right)^{2} \leq 4$ $j_0^{*}$ seems to be proportional to the equilibrium mobility as the prefactor in the exponential function of $j_0^{*}$ is found to be $0.45$, very close to the predicted value of $0.5$~\cite{nikitenko2007a, arkhipov2001a} in case of a constant transport energy $E_{trans}=0$. $\tau$ and $\beta$ have very similar prefactors, but inverse temperature scaling. However, the physical origin of this scaling might be the same, as both parameters define the dynamics of the relaxation. Thus, it is very likely that they are related to the average charge carrier trapping rates limiting relaxation.~\cite{arkhipov1982, nikitenko2007a} 

%\bibliography{MYScottPaper}

%merlin.mbs apsrev4-1.bst 2010-07-25 4.21a (PWD, AO, DPC) hacked
%Control: key (0)
%Control: author (8) initials jnrlst
%Control: editor formatted (1) identically to author
%Control: production of article title (-1) disabled
%Control: page (0) single
%Control: year (1) truncated
%Control: production of eprint (0) enabled
%

\end{document}